# Evaluation of coefficient of friction in bulk metal forming[*]

Soheil Solhjoo[†]

Faculty of Mathematics and Natural Sciences, University of Groningen, Groningen, The Netherlands

**Abstract**

In this study an upper bound analysis for cylindrical "Barrel Compression Test" (BCT) is developed. BCT method is a very simple method which can be utilized in order to evaluate quantitatively the coefficient of friction by means of just one cylindrical specimen in an upsetting test. The method is checked by a series of finite element method (FEM) simulations and by means of the results of FEM simulations the method is modified.

**Keywords:** Barrel Compression Test; bulk deformation; coefficient of friction; upper bound method; finite element simulation.

## 1-Introduction:

The most accepted ways of characterizing friction quantitatively are to define a coefficient of friction or a friction factor at the die/workpiece interface. There are many researchers studied on the interface friction and readers may find some good introductory in references [1-2].

---

[*] Written in summer of 2010, at Department of Materials Science and Engineering, Sharif University of Technology, Tehran, Iran.

[†] soheilsolhjoo@yahoo.com

| **Nomenclature** | |
| --- | --- |
| $b$ | the barrel parameter |
| $H_0$ | initial height of cylinder |
| $H_1$ | height of cylinder after deformation |
| $\Delta H$ | reduction of height of cylinder after deformation |
| $k, \bar{k}$ | current and mean shear yield stress of material, respectively |
| $m$ | constant friction factor |
| $\bar{P}$ | average external pressure applied to cylinder in compression |
| $R, \theta, y$ | general cylindrical coordinate |
| $R_0$ | initial radius of cylinder |
| $R_M, R_T$ | maximum and top radius of cylinder after deformation |
| $\bar{R}$ | average radius of cylinder after deformation |
| $\Delta R$ | difference between maximum and top radius |
| $\sum R$ | summation of maximum and top radius |
| $S_D$ | velocity discontinuity surface |
| $S_F$ | friction surface |
| $\dot{U}$ | die velocity |
| $\dot{U}_\theta, \dot{U}_R, \dot{U}_y$ | velocity components in cylindrical coordinate |
| $V$ | volume of deformation zone |
| $\lvert \Delta v_S \rvert$ | magnitude of sliding velocity on $S_F$ |
| $\lvert \Delta v_t \rvert$ | magnitude of velocity discontinuity tangent to $S_D$ |
| $\dot{W}_t$ | upper bound applied power |
| $\dot{W}_i, \dot{W}_f, \dot{W}_s$ | power dissipation due to internal deformation, friction, internal velocity discontinuity and external traction force, respectively |
| $Y$ | flow stress of material |
| $\dot{\varepsilon}_{ij}$ | components of strain rate tensor |
| $\bar{\dot{\varepsilon}}$ | equivalent strain rate |
| $\mu$ | Coulomb coefficient of friction |
| $\mu_c$ | calculated coefficient of friction from the simulations |
| $\tau$ | shear stress |

The upsetting at room temperature is one of the most widely used workability tests. As the sample is compressed in the presence of friction, it tends to barrel. Variation of the friction conditions and of the upset cylinder's aspect ratio makes changes on the barrel curvature. Avitzur [3] analyzed the barrel compression test (BCT) by means of upper bound theorem and found a relationship between *b* (which was introduced as an arbitrary coefficient) and the friction factor. Ebrahimi and Najafizadeh [1] proposed a method in order to calculate the value of *b* and showed that there is a relationship between *b* and barrel curvature. Very recently, Solhjoo [4] utilized their method and showed that the results of this method need some modifications.

Usually the calculations of metal forming analyses facilitate using the friction factor but most of the computer simulation programs use the coefficient of friction. Therefore, it is important to determine the coefficient of friction of interfaces precisely in order to have a reliable simulation. In this study using the upper bound theorem, the BCT method is analyzed and a relationship is derived which can be used to determine the coefficient of friction. Afterwards, a series of finite element method (FEM) simulations are done in order to check the derived formula. Using the results, the model is modified. The major advantage of the BCT method is that it involves only the physical measurement of the changes in shape.

## 2-Coulomb coefficient of friction

A relative motion between two bodies in contact provides a resistance to this motion which is called friction. The surface area of contact is a boundary of the deformed metal. Thus, the friction resistance is also the shear stress in the material at its boundary. If the friction is assumed to obey coulomb's law, then:

$$\tau = \mu P \tag{1}$$

The shear stress at any point on that surface is proportional to the pressure (*P*) between the two bodies and is directed opposite to relative motion between those bodies. Since the value of *P* is different at any point, the value of coefficient of friction would be different for each point. As a result, any calculations will be far too complex. In order to solve this problem an average value for the pressure can be defined which leads to a single mean value for coefficient of friction and also taken as a constant for a given die and material (under constant surface and temperature conditions). This value is also considered independent of velocity [3].

## 3-Evaluation of coefficient of friction

The value of coefficient of friction could be obtained by two different procedures. First one is to calculate the value of coefficient of friction from the constant friction factor. This method starts from two different formulae for shear stress ($\tau = \mu \bar{P} = mk$). In BCT the value of $\bar{P}$ evaluates as follows:

$$\bar{P} = Y\left(1 + \frac{2m}{3\sqrt{3}} \frac{\bar{R}}{H_1}\right) \tag{2}$$

where $\bar{R} = R_0 \sqrt{\frac{H_0}{H_1}}$ [1] determined from volume constancy. Using von Mises' yield criterion one can find $Y = \sqrt{3}k$ which leads to:

$$\frac{\bar{P}}{k} = \frac{m}{\mu} = \sqrt{3} + \frac{2m}{3} \frac{\bar{R}}{H_1} \tag{3}$$

By means of this equation the value of coefficient of friction can be easily found from constant friction factor.

$$\mu = \frac{3m}{3\sqrt{3} + 2m\left(\frac{\bar{R}}{H_1}\right)} \tag{4}$$

Additionally, it is possible to analyze BCT by means of the upper bound theorem in order to determine the value of the coefficient of friction directly from the test results without undue calculations of constant friction factor.

# 4-Upper bound method

The upper bound method (UBM) is based on the energy principle known as the *upper bound theorem* [5]. The upper bound theorem states that the rate of total energy associated with any kinematically admissible velocity field defines an upper bound to the actual rate of total energy required for the deformation. Hence, for a given class of kinematically admissible velocity fields the velocity field that minimizes the rate of total energy is the lowest upper bound and therefore is nearest to the actual solution.

The upper bound theorem was formulated by Prager and Hodge [6] and later modified by Drucker et al. [7-9] by including the velocity discontinuities. Kudo [10] introduced the concept of dividing the workpiece into several rigid blocks, obtaining lower upper bounds by changing the shape and number of these blocks. Kobayashi and Thomsen [11] suggested curved discontinuity surfaces for the deformation blocks which gave better upper bounds for some axisymmetric problems. The upper bound theorem states that the actual energy rate is less than or equal to:

$$\dot{W}_t = \dot{W}_i + \dot{W}_f + \dot{W}_s \tag{5a}$$

$$\dot{W}_i = 2\int_V \bar{k}\bar{\dot{\varepsilon}}dV \tag{5b}$$

$$\dot{W}_f = \int_{S_F} \tau |\Delta v_s| dS \tag{5c}$$

$$\dot{W}_s = \int_{S_D} k |\Delta v_t| dS \tag{5d}$$

where $\bar{\dot{\varepsilon}}$ can be determined by Eq.6:

$$\bar{\dot{\varepsilon}} = \sqrt{\tfrac{1}{2}\dot{\varepsilon}_{ij}\dot{\varepsilon}_{ij}} \tag{6}$$

The third term of Eq.5a ($\dot{W}_S$), also known as the *jump condition,* can be omitted when a class of continuous velocity fields is considered [5].

## 5-Analysis of BCT

The coordinates system applied to a solid disk is plainly illustrated in Fig.1. Two anvils move toward each other at the same absolute velocity. Since the radial component of velocity ($\dot{U}_R$) in the center of the disk (y=0) is larger than $\dot{U}_R$ at the friction surfaces (surfaces in contact with anvils i.e. y=±H$_1$/2), the cylinder barrels out during the compression.

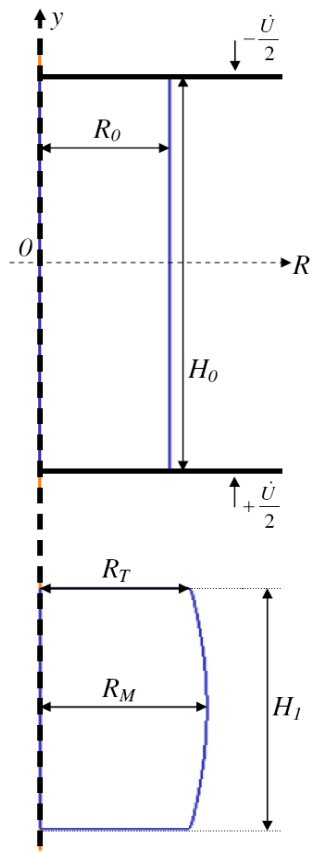

**Fig.1.** A simple representation of barrel compression test. (upper) coordinates system, (lower) the disk after upsetting.

As Avitzur clarified [3] friction reduces velocity components on the platen surfaces which lead to reduced friction loss. But since $\dot{U}_R$ changes over the thickness of the billet, a shear component introduces which increases the internal power of deformation. Because of this, the velocity at the die surface may decreases but does not drop to zero. To sum up, bulging produces a slightly lower total energy.

In order to make the paper easier to read, the whole mathematical steps taken for the analysis of BCT by means of UBM is written in appendix A. The analysis results in the following equation which can be used in order to evaluate the value of coefficient of friction.

$$\mu = \frac{3b\dfrac{\bar{R}}{H_1}}{12 - 2b\left[1 - \left(\dfrac{\bar{R}}{H_1}\right)^2\right]} \quad (7)$$

where the value of $b$ defines as [4]:

$$b = 4\frac{\Delta R}{\Sigma R}\left(2\frac{H_1}{\Delta H} - 1\right) \quad (8)$$

# 6-Examination of BCT method

Finite element method (FEM) is employed to check the correctness of BCT method.

*6.1-Finite element method*

The FEM was conducted using DEFORM-2D V9.0 software (Scientific Forming Technologies Corp.) and a series of simulations of upsetting tests were performed. Due to the axisymmetric profile of the simulation, only one half of each cylindrical billet and die was modeled. The value of $H_0$ and $R_0$ were selected as 16 and 5mm, respectively. - Using the materials library of the software, aluminum (Al-6063) was selected as the material of the billet. Rigid bodies were suggested for the upper and lower platens to reduce the running time during the simulations.

*6.2-Simulation control of upsetting process*

In these simulations, the coordinates system's perpendicular direction was set as *y*-axis. During the upsetting process, the upper die, which was primary die, with a given speed moved along the negative *y*-axis and the billet was placed between the moving upper and the still lower platens.

*6.3-Simulation procedures*

Different final heights (between 15 and 12mm with 1mm steps) and different coefficients of friction (0.01, 0.03, 0.05, 0.07, 0.1, 0.2, 0.3, 0.4, 0.5 and 0.577) were selected for simulations.

# 7-Results and discussion

After each simulation the values of $R_M$ and $R_T$ were measured and consequently the values of parameter *b* and coefficient of friction were calculated. The values of $R_M$, $R_T$, parameter *b* and $\mu$ are listed in table 1.

**Table 1.** The measured and calculated values of $R_M$, $R_T$, $b$ and $\mu_c$ from the simulations performed under different constant friction factors with different final heights.

| μ | 0.01 | 0.03 | 0.05 | 0.07 | 0.1 | 0.2 | 0.3 | 0.4 | 0.5 | 0.577 |
|---|---|---|---|---|---|---|---|---|---|---|
| $H_1$ | 15 | | | | | | | | | |
| $R_M$ | 5.166 | 5.173 | 5.180 | 5.186 | 5.195 | 5.220 | 5.229 | 5.230 | 5.230 | 5.230 |
| $R_T$ | 5.156 | 5.143 | 5.129 | 5.117 | 5.100 | 5.052 | 5.027 | 5.021 | 5.020 | 5.020 |
| $B$ | 0.112 | 0.337 | 0.574 | 0.777 | 1.070 | 1.897 | 2.285 | 2.365 | 2.377 | 2.377 |
| $\mu_c$ | 0.010 | 0.031 | 0.054 | 0.075 | 0.109 | 0.226 | 0.296 | 0.312 | 0.314 | 0.314 |
| $H_1$ | 14 | | | | | | | | | |
| $R_M$ | 5.347 | 5.355 | 5.363 | 5.37 | 5.382 | 5.418 | 5.437 | 5.44 | 5.44 | 5.441 |
| $R_T$ | 5.333 | 5.312 | 5.292 | 5.271 | 5.241 | 5.144 | 5.082 | 5.073 | 5.069 | 5.066 |
| $B$ | 0.068 | 0.210 | 0.347 | 0.484 | 0.690 | 1.349 | 1.755 | 1.815 | 1.836 | 1.856 |
| $\mu_c$ | 0.007 | 0.021 | 0.035 | 0.050 | 0.073 | 0.159 | 0.223 | 0.234 | 0.237 | 0.241 |
| $H_1$ | 13 | | | | | | | | | |
| $R_M$ | 5.550 | 5.562 | 5.574 | 5.586 | 5.602 | 5.651 | 5.671 | 5.674 | 5.675 | 5.675 |
| $R_T$ | 5.528 | 5.497 | 5.465 | 5.434 | 5.388 | 5.235 | 5.190 | 5.179 | 5.173 | 5.170 |
| $B$ | 0.061 | 0.180 | 0.303 | 0.423 | 0.597 | 1.172 | 1.358 | 1.399 | 1.419 | 1.428 |
| $\mu_c$ | 0.007 | 0.020 | 0.034 | 0.048 | 0.069 | 0.149 | 0.178 | 0.184 | 0.188 | 0.189 |
| $H_1$ | 12 | | | | | | | | | |
| $R_M$ | 5.777 | 5.795 | 5.813 | 5.829 | 5.853 | 5.918 | 5.938 | 5.941 | 5.943 | 5.943 |
| $R_T$ | 5.747 | 5.703 | 5.659 | 5.615 | 5.548 | 5.345 | 5.300 | 5.289 | 5.276 | 5.271 |
| $B$ | 0.052 | 0.160 | 0.268 | 0.374 | 0.535 | 1.017 | 1.135 | 1.161 | 1.189 | 1.199 |
| $\mu_c$ | 0.006 | 0.020 | 0.033 | 0.047 | 0.069 | 0.141 | 0.160 | 0.164 | 0.169 | 0.170 |

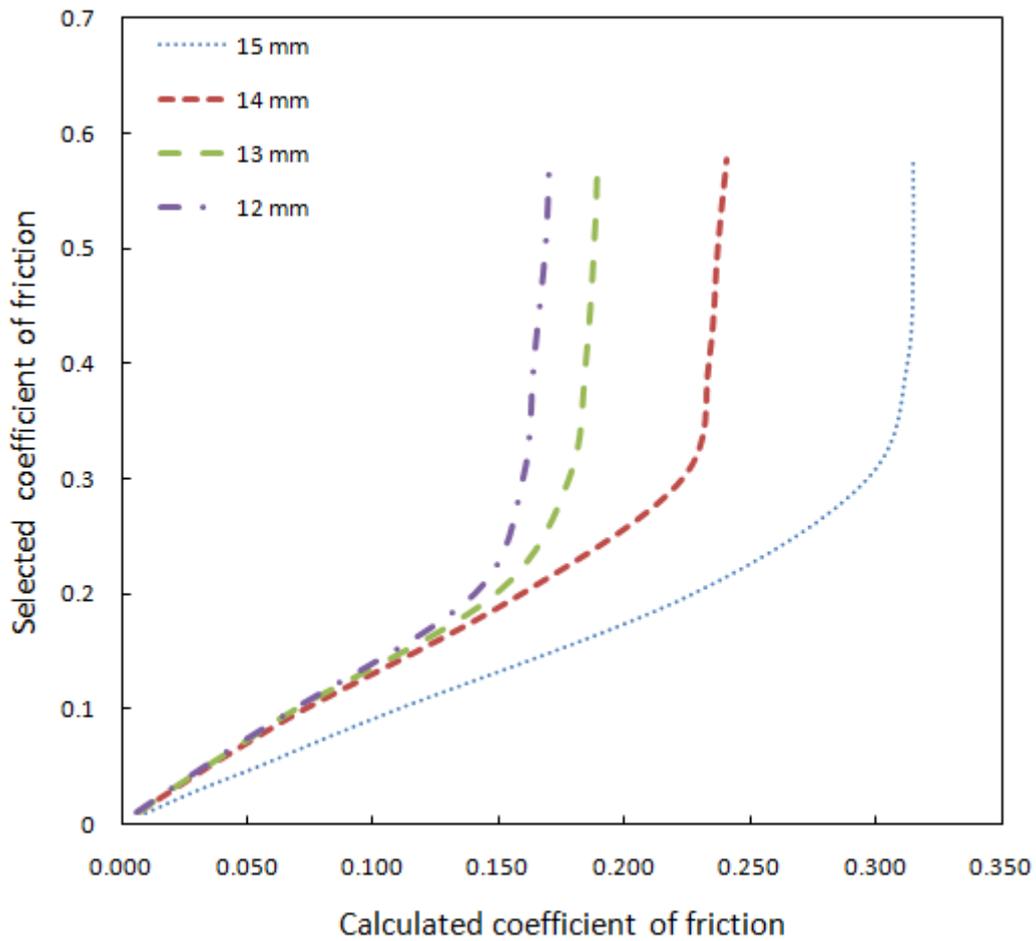

**Fig.2.** Values of $\mu$ vs. $\mu_c$.

As Fig.2 indicates, the values of selected $\mu$ for simulations and $\mu_c$ (calculated $\mu$ from simulations) do not match. This miscalculation is the result of the presumption that parameter *b* has a very small value which leads to omission of all squared and greater terms of *b* in the procedure of UBM. It should be noted that UBM could not be carried out without such assumption. However, this problem could be overcome with some modifications. For this reason, Fig.2 can be used as a friction calibration curve (FCC). Furthermore, since the values of $\mu$ depend on aspect ratio of the billet and also the percentage of height reduction, BCT method can be used to generate FCCs. Besides, the value of $\mu$ as a function of $\mu_c$ can be expressed by the following equation:

$$\mu = u_1 \exp(u_2 \mu_c) + u_3 \mu_c \qquad (9)$$

where $u_1$, $u_2$ and $u_3$ are constants which have different values for different reduction in heights. The calculated values for these three constants based on the applied simulations are listed in table 2. It is considerable in Fig.2, lower reduction in height results in higher accuracy of calculations and also acquires a wider range for $\mu_c$ that make it easier to use the FCCs.

**Table 2.** The values of $u_1$, $u_2$ and $u_3$ at different final heights

| $H_1$ (mm) | %Reduction in height | $u_1 (\times 10^{-8})$ | $u_2$ | $u_3$ |
|---|---|---|---|---|
| 15 | 6.25 | 5.917 | 48.28 | 0.8536 |
| 14 | 12.5 | $6.811 \times 10^{-2}$ | 81.95 | 1.235 |
| 13 | 18.75 | 6.548 | 80.88 | 1.282 |
| 12 | 25 | $7.704 \times 10^{-1}$ | 103.2 | 1.351 |

## 8-Conclusion

In this paper the barrel compression test is analyzed by means of UBM and a formula is derived for evaluation of coefficient of friction quantitatively from BCT method. The method was tested by a series of FEM simulations (for Al-6063) which showed that this method needs to be modified. This modification can be done using FCCs or Eq.9. Also, it showed that this method can be used to generate FCCs in order to determine the coefficient of friction at the die/workpiece interface in large deformation processes. In addition, it is noted that upsetting under low percentage of reduction in height yields more reliable and accurate results.

# 9-Appendix A

The axes of a cylindrical coordination system are the radial direction R, the circumferential direction $\vartheta$ and the direction of the axis of symmetry y. The velocity is $\dot{U}(\dot{U}_R, \dot{U}_\theta, \dot{U}_y)$ while the strain rate components $\dot{\varepsilon}_{ij}$ acquire the subscripts R, $\vartheta$ and y. The strain rates as functions of the velocity components are:

$$\dot{\varepsilon}_{RR} = \frac{\partial \dot{U}_R}{\partial R} \qquad (A-1a)$$

$$\dot{\varepsilon}_{\theta\theta} = \frac{\dot{U}_R}{R} + \frac{1}{R}\frac{\partial \dot{U}_\theta}{\partial \theta} \qquad (A-1b)$$

$$\dot{\varepsilon}_{yy} = \frac{\partial \dot{U}_y}{\partial y} \qquad (A-1c)$$

$$\dot{\varepsilon}_{R\theta} = \frac{1}{2}\left(\frac{1}{R}\frac{\partial \dot{U}_R}{\partial \theta} + \frac{\partial \dot{U}_\theta}{\partial R} - \frac{\dot{U}_\theta}{R}\right) \qquad (A-1d)$$

$$\dot{\varepsilon}_{\theta y} = \frac{1}{2}\left(\frac{\partial \dot{U}_\theta}{\partial y} + \frac{1}{R}\frac{\partial \dot{U}_y}{\partial \theta}\right) \qquad (A-1e)$$

$$\dot{\varepsilon}_{yR} = \frac{1}{2}\left(\frac{\partial \dot{U}_R}{\partial y} + \frac{\partial \dot{U}_y}{\partial R}\right) \qquad (A-1f)$$

The axes were chosen as shown in Fig.1. The origin of the cylindrical coordinate system is at the center of the disk. The two platens are considered rigid bodies and move toward each other at the same absolute velocity ($\frac{\dot{U}}{2}$). Because of symmetry and to ease computation, the upper half of the disk is considered. Assume a velocity field [3] for $0 \leq y \leq \frac{H_1}{2}$ only

$$\dot{U}_\theta = 0 \qquad (A-2a)$$

$$\dot{U}_R = A\dot{U}\frac{R}{H_1}\exp\left(-\frac{by}{H_1}\right) \qquad (A-2b)$$

$$\dot{U}_y = \dot{U}_y(R, y) \qquad (A-2c)$$

where *b* is the barrel parameter and determines the amount of bulge. Eq.(*A-2a*) is on the basis of this assumption that no rotation of the disk occurs in the course of the deformation. The strain rate field, by Eqs.(*A-1a*)-(*A-1f*), becomes:

$$\dot{\varepsilon}_{RR} = \frac{\partial \dot{U}_R}{\partial R} = A\frac{\dot{U}}{H_1}\exp\left(-\frac{by}{H_1}\right) \qquad (A-3a)$$

$$\dot{\varepsilon}_{\theta\theta} = \frac{\partial \dot{U}_R}{\partial R} = \dot{\varepsilon}_{RR} \qquad (A-3b)$$

$$\dot{\varepsilon}_{yy} = \frac{\partial \dot{U}_y}{\partial y} \qquad (A-3c)$$

Because of volume constancy:

$$\dot{\varepsilon}_{RR} + \dot{\varepsilon}_{\theta\theta} + \dot{\varepsilon}_{yy} = 0 \rightarrow \frac{\partial \dot{U}_y}{\partial y} = -2A\frac{\dot{U}}{H_1}\exp\left(-\frac{by}{H_1}\right) \qquad (A-4)$$

Integration of Eq.(*A-4*) shows that:

$$\dot{U}_y = 2\frac{A}{b}\frac{\dot{U}}{H_1}\exp\left(-\frac{by}{H_1}\right) + f_{(R)} \qquad (A-5a)$$

$$f_{(R)} = \left[\dot{U}_y - 2\frac{A}{b}\dot{U}\exp\left(-\frac{by}{H_1}\right)\right]\bigg|_{y=\frac{H_1}{2}} = -\frac{\dot{U}}{2} - 2\frac{A}{b}\dot{U}\exp\left(-\frac{b}{2}\right) \qquad (A-5b)$$

The velocity field of height component becomes:

$$\dot{U}_y = -\frac{\dot{U}}{2}\left[1 + 4\frac{A}{b}\left(\exp\left(-\frac{b}{2}\right) - \exp\left(-\frac{by}{H_1}\right)\right)\right] \qquad (A-6)$$

On account of symmetry:

$$\dot{U}_y\Big|_{y=0} = 0 = -\frac{\dot{U}}{2}\left[1 + 4\frac{A}{b}\left(\exp\left(-\frac{b}{2}\right) - \exp\left(-\frac{by}{H_1}\right)\right)\right] \qquad (A-7a)$$

which evaluates $A$ to be:

$$A = \frac{1}{4}\frac{-b}{\exp\left(-\frac{b}{2}\right) - 1} \qquad (A-7b)$$

Therefore, the velocity field becomes:

$$\dot{U}_\theta = 0 \qquad (A-8a)$$

$$\dot{U}_R = \frac{1}{4}\frac{-b}{\exp\left(-\frac{b}{2}\right) - 1}\dot{U}\frac{R}{H_1}\exp\left(-\frac{by}{H_1}\right) \qquad (A-8b)$$

$$\dot{U}_y = -\frac{\dot{U}}{2}\left[1 - \frac{\exp\left(-\frac{b}{2}\right) - \exp\left(-\frac{by}{H_1}\right)}{\exp\left(-\frac{b}{2}\right) - 1}\right] = -\frac{\dot{U}}{2}\frac{1 - \exp\left(-\frac{by}{H_1}\right)}{1 - \exp\left(-\frac{b}{2}\right)} \qquad (A-8c)$$

$$\dot{\varepsilon}_{RR} = \dot{\varepsilon}_{\theta\theta} = -\frac{1}{2}\dot{\varepsilon}_{yy} = \frac{1}{4}\frac{-b}{\exp\left(-\frac{b}{2}\right) - 1}\frac{\dot{U}}{H_1}\exp\left(-\frac{by}{H_1}\right) \qquad (A-9a)$$

$$\dot{\varepsilon}_{Ry} = \frac{1}{2}\left(\frac{\partial \dot{U}_R}{\partial y} + \frac{\partial \dot{U}_y}{\partial R}\right) = \frac{1}{8}\frac{b^2}{\exp\left(-\frac{b}{2}\right) - 1}\dot{U}\frac{R}{H_1^2}\exp\left(-\frac{by}{H_1}\right) \qquad (A-9b)$$

$$\dot{\varepsilon}_{R\theta} = \dot{\varepsilon}_{\theta y} = 0 \qquad (A-9c)$$

By using Eqs.(*A-9a*)-(*A-9c*) the internal power of deformation becomes:

$$\dot{W}_i = \frac{2}{\sqrt{3}} Y \int_V \sqrt{\frac{1}{2} \dot{\varepsilon}_{ij} \dot{\varepsilon}_{ij}} dV = \frac{2}{\sqrt{3}} Y \int_V \sqrt{3\dot{\varepsilon}_{RR}^2 + \dot{\varepsilon}_{Ry}^2} dV \qquad (A-10a)$$

where $dV = 2\pi R\, dR\, dy$. Therefore:

$$\dot{W}_i = \frac{2}{\sqrt{3}} Y 2\pi \frac{2\dot{U}}{4T} \frac{-b}{\exp\left(-\frac{b}{2}\right)-1} \times \int_{y=o}^{y=\frac{H_1}{2}} \left[ \exp\left(-\frac{by}{H_1}\right) \left( \int_{R=0}^{\bar{R}} R\sqrt{3 + \frac{1}{4}b^2 \frac{R^2}{H_1^2}}\, dR \right) \right] dy \qquad (A-10b)$$

$$\dot{W}_i = -\frac{Y}{3\sqrt{3}} \pi \frac{\dot{U}}{T} b\bar{R}^3 \left[ \left(1 + \frac{12}{\bar{R}^2} \frac{H_1^2}{b^2}\right)^{\frac{3}{2}} - \left(\frac{12}{\bar{R}^2} \frac{H_1^2}{b^2}\right)^{\frac{3}{2}} \right] \qquad (A-10c)$$

Minimum $\dot{W}_i$ is obtained for *b=0* and it represents the ideal power of deformation where there is no bulging in the specimen. But in actual solution $b \neq 0$ and the internal power of deformation as computed by Eq.(*A-10c*) is higher than the ideal value.

The friction loss is:

$$\dot{W}_f = -4\pi \mu \bar{P} \int_{R=0}^{\bar{R}} \dot{U}_R \bigg|_{y=\frac{H_1}{2}} R\, dR = \pi \mu \bar{P} b \frac{\exp\left(-\frac{b}{2}\right)}{\exp\left(-\frac{b}{2}\right)-1} \frac{\dot{U}}{H_1} \int_{R=0}^{\bar{R}} R^2 dR \qquad (A-11a)$$

Multiplying the numerator and denominator of Eq.(*A-11a*) by $\exp\left(\frac{b}{2}\right)$ results in:

$$\dot{W}_f = \frac{\bar{P}}{3} \mu \pi \frac{\dot{U}}{T} b\bar{R}^3 \frac{1}{1-\exp\left(\frac{b}{2}\right)} \qquad (A-11b)$$

The external power is supplied by two platens and equals:

$$\dot{W}_t = -\pi \overline{R}^2 \overline{P} \cdot 2 \frac{\dot{U}}{2} = -\pi \overline{R}^2 \dot{U} \overline{P} \qquad (A-12)$$

Equating the external power with the internal power of deformation plus friction loss, one obtains:

$$\frac{\overline{P}}{Y} = -\frac{\dot{W}_t}{\pi \overline{R}^2 \dot{U} Y}$$

$$= \frac{1}{3\sqrt{3}} b \frac{\overline{R}}{H_1} \left[ \left(1 + \frac{12}{\overline{R}^2} \frac{H_1^2}{b^2}\right)^{\frac{3}{2}} - \left(\frac{12}{\overline{R}^2} \frac{H_1^2}{b^2}\right)^{\frac{3}{2}} \right] - \frac{1}{3} \mu \frac{\overline{P}}{Y} b \frac{\overline{R}}{H_1} \frac{1}{1 - \exp\left(\frac{b}{2}\right)} \qquad (A-13a)$$

This can be simplified to:

$$\frac{\overline{P}}{Y} = \frac{\frac{1}{3\sqrt{3}} b \frac{\overline{R}}{H_1} \left[ \left(1 + \frac{12}{\overline{R}^2} \frac{H_1^2}{b^2}\right)^{\frac{3}{2}} - \left(\frac{12}{\overline{R}^2} \frac{H_1^2}{b^2}\right)^{\frac{3}{2}} \right]}{1 - \frac{1}{3} \mu b \frac{\overline{R}}{H_1} \frac{1}{\exp\left(\frac{b}{2}\right) - 1}} = \frac{S}{M} \qquad (A-13b)$$

where *S* and *M* stand as numerator and denominator of Eq.(*A-13b*). *S* can be rearrange as follows:

$$S = 8 \frac{\overline{R}}{H_1} b \left[ \left(\frac{1}{12} + \left(\frac{H_1}{\overline{R}}\right)^2 \frac{1}{b^2}\right)^{\frac{3}{2}} - \left(\frac{H_1}{\overline{R}}\right)^3 \frac{1}{b^3} \right] = 8 \left(\frac{H_1}{\overline{R}}\right)^2 \frac{1}{b^2} \left[ \left(1 + \frac{1}{12} \left(\frac{\overline{R}}{H_1}\right)^2 b^2\right)^{\frac{3}{2}} - 1 \right] \qquad (A-14)$$

Since the actual shape of disk is such that the required power is minimized, to have minimum value of $\overline{P}$ the optimum *b* must be chosen. The value minimizing the pressure can be determined by successive approximations directly from the above equation. Differentiating the above equation with respect to *b* and equating the derivative to zero results in an implicit equation which offers no advantage over Eq.(*A-14*). But some advantage can be gained by substituting the proper series expansions as follows. First let's define two variables as follows:

$$C = \frac{1}{12}\left(\frac{\overline{R}}{H_1}\right)^2 b^2 \Rightarrow \frac{\partial C}{\partial b} = \frac{1}{6}\left(\frac{\overline{R}}{H_1}\right)^2 b \quad (A-15a)$$

$$\beta = \frac{b}{2} \Rightarrow \frac{\partial \beta}{\partial b} = \frac{1}{2} \quad (A-15b)$$

Therefore, *S* and *M* become:

$$S = \frac{2}{3}\frac{1}{C}\left[(1+C)^{\frac{3}{2}} - 1\right] \quad (A-16a)$$

$$M = 1 - \frac{2}{3}\mu\frac{\overline{R}}{H_1}\frac{\frac{b}{2}}{\exp\left(\frac{b}{2}\right) - 1} = 1 - \frac{2}{3}\mu\frac{\overline{R}}{H_1}\frac{\beta}{\exp(\beta) - 1} \quad (A-16b)$$

In order to solve Eqs.(*A-16a*) and (*A-16b*) two following series expansions are needed:

$$(1+C)^{\frac{3}{2}} = 1 + \frac{3}{2}C + \frac{3}{2}\frac{1}{4}C^2 - \frac{3}{2}\frac{1}{4}\frac{1}{6}C^3 + \frac{3}{2}\frac{1}{4}\frac{1}{6}\frac{3}{8}C^4 - \frac{3}{2}\frac{1}{4}\frac{1}{6}\frac{3}{8}\frac{5}{10}C^5 + \ldots \quad (A-17a)$$

$$\frac{\beta}{\exp(\beta) - 1} = \sum_{n=0}^{\infty}\frac{B_n \beta^n}{n!} \quad (A-17b)$$

where $B_n$'s are called Bernoulli's numbers and sometimes called *even-index* Bernoulli numbers. The first few Bernoulli numbers ($B_n$) are $B_0 = 1$, $B_1 = -\frac{1}{2}$, $B_2 = \frac{1}{6}$, $B_4 = -\frac{1}{30}$, $B_6 = \frac{1}{42}$ and $B_8 = -\frac{1}{30}$. Expanding the series, *S* and *M* become:

$$S = 1 + \frac{1}{4}C - \frac{1}{4}\frac{1}{6}C^2 + \frac{1}{4}\frac{1}{6}\frac{3}{8}C^3 - \frac{1}{4}\frac{1}{6}\frac{3}{8}\frac{5}{10}C^4 + \ldots \qquad (A-18a)$$

$$M = 1 - \frac{2}{3}\mu\frac{\overline{R}}{H_1}\left(1 - \frac{\beta}{2} + \frac{1}{6}\frac{\beta^2}{2!} - \frac{1}{30}\frac{\beta^4}{4!} + \frac{1}{42}\frac{\beta^6}{6!} - \frac{1}{30}\frac{\beta^8}{8!} + \ldots\right) \qquad (A-18b)$$

The expression $\dfrac{\overline{P}}{Y} = \dfrac{S}{M}$ can be differentiated with respect to *b* and the derivative can be equated to zero for the optimum value of *b*.

$$\frac{\partial}{\partial b}\left(\frac{\overline{P}}{Y}\right) = \frac{M\dfrac{\partial S}{\partial b} - S\dfrac{\partial M}{\partial b}}{M^2} = 0 \Rightarrow M\frac{\partial S}{\partial b} = S\frac{\partial M}{\partial b} \qquad (A-19)$$

$$\frac{\partial S}{\partial b} = \left(\frac{1}{4} - \frac{1}{4}\frac{1}{6}2C + \frac{1}{4}\frac{1}{6}\frac{3}{8}3C^2 + \ldots\right)\frac{\partial C}{\partial b}$$

$$= \left[\frac{1}{4} - \frac{1}{144}\left(\frac{\overline{R}}{H_1}\right)^2 b^2 + \frac{3}{64}\frac{1}{144}\left(\frac{\overline{R}}{H_1}\right)^4 b^4 + \ldots\right]\frac{1}{6}\left(\frac{\overline{R}}{H_1}\right)^2 b \qquad (A-20a)$$

$$\frac{\partial M}{\partial b} = -\frac{2}{3}\mu\frac{\overline{R}}{H_1}\left(-\frac{1}{2} + \frac{1}{6}\frac{2\beta}{2!} - \frac{1}{30}\frac{4\beta^3}{4!} + \ldots\right)\frac{\partial \beta}{\partial b} = -\frac{1}{3}\mu\frac{\overline{R}}{H_1}\left(-\frac{1}{2} + \frac{b}{12} - \frac{b^3}{1440} + \ldots\right) \qquad (A-20b)$$

Substituting the expansions in $M\dfrac{\partial S}{\partial b} = S\dfrac{\partial M}{\partial b}$ and omitting all squared and greater terms of *b*:

$$\left[-\frac{1}{36}\mu\left(\frac{\overline{R}}{H_1}\right)^3 + \frac{1}{24}\left(\frac{\overline{R}}{H_1}\right)^2 + \frac{1}{36}\mu\left(\frac{\overline{R}}{H_1}\right)\right]b - \frac{1}{6}\mu\left(\frac{\overline{R}}{H_1}\right) = 0 \qquad (A-21)$$

This equation can be solved explicitly for the optimum *b*:

$$b = \frac{12\mu\dfrac{H_1}{\overline{R}}}{3 + 2\mu\left(\dfrac{H_1}{\overline{R}} - \dfrac{\overline{R}}{H_1}\right)} \qquad (A-22)$$

which with a rearrangement can be rewritten as follows:

$$\mu = \frac{3b}{12\frac{H_1}{\bar{R}} - 2b\left(\frac{H_1}{\bar{R}} - \frac{\bar{R}}{H_1}\right)} = \frac{3b\frac{\bar{R}}{H_1}}{12 - 2b\left[1 - \left(\frac{\bar{R}}{H_1}\right)^2\right]} \qquad (A-23)$$